\documentclass[runningheads]{llncs}

\usepackage[T1]{fontenc}
\usepackage{graphicx}

\usepackage{mathtools} 
\usepackage{amsmath}  
\usepackage{amssymb}
\usepackage{amsfonts}
\usepackage{bm}
\usepackage{booktabs}       
\usepackage{array}          
\usepackage{threeparttable} 
\usepackage{adjustbox}      
\usepackage{caption}        
\usepackage{multirow}       
\usepackage{makecell}       
\usepackage{calc}           
\usepackage{wrapfig}  
\usepackage{comment}
\usepackage{hyperref}
\newcolumntype{L}{l}

\begin{document}

\title{PhaseGAN: High-Fidelity Vocoder via Decoupled Amplitude and GAN-Driven Phase Reconstruction}
\titlerunning{PhaseGAN}

\author{Wenzheng Zhang\inst{1} \and
Xueliang Zhang\inst{1}\thanks{Corresponding author.} \and
Shulin He\inst{1} \and
Fei Zhao\inst{1} \and
Xin Liu\inst{1} \and
Pengjie Shen\inst{1} \and
Zhenlong Guo\inst{1} \and
Zixuan Xue\inst{1} \and
Hongtao Bao\inst{1} \and
Zixuan Li\inst{1}}
\authorrunning{W. Zhang et al.}
\institute{College of Computer Science, Inner Mongolia University, China}

\maketitle
\begin{abstract}
A vocoder is a pivotal component of modern text-to-speech (TTS) systems. Despite the significant progress of neural network-based vocoders, accurate phase reconstruction remains the main challenge limiting both audio quality and modeling efficiency. We introduce PhaseGAN, a lightweight vocoder that addresses this limitation through a "mel → Amplitude → Phase" reconstruction pipeline. By reconstructing amplitude and phase spectra via distinct methodologies, the proposed PhaseGAN outperforms state-of-the-art baselines while utilizing fewer model parameters and reduced computational requirements. The compact version generates high-fidelity audio with approximately 500K parameters and 1 GMAC computational load, making it highly suitable for real-time applications on edge devices. In addition, our approach exhibits exceptional musical audio synthesis capabilities despite no training on musical data, illustrating unprecedented cross-domain generalization. See \url{https://github.com/phasegan/phasegan-audio-demo} for demos of our work.

\end{abstract}
\section{Introduction}
Speech and audio synthesis \cite{1,2} have undergone transformative advances with the advent of deep learning, enabling diverse applications from human-computer interaction to creative arts \cite{fastspeech,tacotron2} . Central to these systems are vocoders—algorithms that synthesize high-fidelity audible waveforms from low-dimensional acoustic representations such as mel-spectrograms.

The performance of a vocoder \cite{4} is evaluated mainly in terms of sound naturalness, intelligibility, resource consumption and synthesis efficiency \cite{3}. Thanks to the rapid development of deep learning, in recent years, neural network-based vocoders (e.g., HiFi-GAN) have far surpassed the early rule-based methods (e.g., Straight \cite{STRAIGHT}) and has dominated this field.

Neural vocoders are typically categorized into five distinct types:
\begin{itemize}
    \item Autoregressive models  (e.g., WaveNet \cite{wavenet}) generate waveforms sample-by-sample. While capable of synthesizing high-fidelity speech, their prohibitively low computational efficiency renders them impractical for real-time applications.
    \item Flow-based models (e.g., WaveGlow \cite{waveglow}) utilize invertible transformations to project speech signals into latent spaces. Although enabling parallel waveform synthesis, these models exhibit two key limitations: excessive parameterization and perceptually inferior audio quality compared with autoregressive vocoders.
    \item GAN \cite{GAN} -based vocoders (e.g., HiFi-GAN \cite{hifigan}) leverage generator-discriminator adversarial training. They achieve real-time synthesis efficiency and competitive quality, yet face training instability issues and spectral artifacts \cite{BIGVGAN}.
    \item Diffusion models (e.g., DiffWave \cite{DiffWave}) reconstruct speech through iterative noise-removal processes. They deliver exceptional stability and perceptual fidelity, but requiring 6-100 sampling iterations.
    \item Variational autoencoders (e.g., FreeCodec \cite{freecodec}) encode speech signals into disentangled latent representations (e.g., timbre, prosody). These compact models ($<$5M parameters) support attribute control, though achieve lower perceptual quality compared with GAN and diffusion counterparts.
\end{itemize}

Compared with other methods, GAN-based vocoders achieve an optimal balance among inference speed \cite{5}, synthesis quality, and versatility, which is suitable for real-world applications like voice assistants, accessibility tools, and emotionally expressive AI systems. While challenges such as training instability and artifacts persist, these approaches remain highly practical solutions \cite{SpectralGANs}.

The essence of a vocoder is a generative task \cite{wavenet}, that is, to generate a sound waveform using low-dimensional acoustic features. From the perspective of the time-frequency domain, the vocoder needs to generate amplitude and phase spectra. If we look at these two parts separately, their generation difficulty is completely different. The reconstruction of the amplitude spectrum is relatively simple, and it is more like an interpolation task, because the low-dimensional acoustic feature input usually contains amplitude information, such as the Mel spectrum. In contrast, the reconstruction of the phase spectrum is more difficult because there is no phase information in the input \cite{phase1,phase2,phase3}. Moreover, the relationship between the amplitude spectrum and the phase spectrum is very complex. One amplitude spectrum corresponds to multiple phase spectra that are exactly the same in auditory perception \cite{DDSP}. Therefore, the reconstruction of the phase spectrum is a generative task. Based on the above thinking, we believe that different methods should be used to reconstruct the amplitude spectrum and the phase spectrum respectively, and this will be helpful for low-resource, efficient and high-fidelity vocoders. We therefore propose a dual-stage vocoder framework: the first stage reconstructs spectral amplitude, while the second generates perceptually coherent phase conditioned on the estimated amplitude spectrum. Compared with the current mainstream T-F domain vocoder methods, we adopted an amplitude-phase decoupling processing approach based on different task difficulties. Moreover, the phase recovery was carried out through generation, and the entire training process of the model did not involve any phase information of the labels.  The main contributions of this work are as follows:
\begin{itemize}
\item For the reconstruction of the amplitude spectrum, we first employ a fixed interpolation method to upscale/transform the Mel-spectrum to the linear frequency scale of the amplitude spectrum. We then reconstruct the amplitude spectrum from this interpolated Mel-spectrum using an Inplace Cepstral Convolutional Recurrent Neural Network (ICCRN) model \cite{iccrn}. The ICCRN model is an inplace convolutional architecture, which abandons down and up sampling operations along frequency axis. ICCRN can not only effectively mitigate the impact of artifacts on high-frequency information, but also significantly reduce the number of parameters and calculations.
\item To address the issue encountered in phase reconstruction—namely, the "one-to-many" mapping in which a single amplitude spectrum can correspond to multiple perceptually similar phase spectra—we depart from strongly supervised approaches such as Apnet2 \cite{apnet2} and FreeV \cite{freeV2}. Instead, we adopt an adversarial training strategy based on the recently proposed R3-GAN \cite{r3gan}, which synthesizes a phase spectrum that closely matches the given amplitude spectrum without relying on any labeled phase information. R3-GAN has demonstrated greater training stability in this setting. To the best of our knowledge, this work represents the first systematic application of such an adversarial training strategy to speech phase reconstruction, thereby establishing a new benchmark for this problem.
\item The proposed method significantly outperforms existing approaches in terms of model size, computational efficiency, and synthesized audio quality. Its design enables deployment on resource-constrained embedded devices. Moreover, the method demonstrates strong generalization on both unseen-speaker test sets and cross-domain Chinese singing data, indicating its potential for broader application scenarios. Experiments on end-to-end speech synthesis further confirm the effectiveness of our approach in this task.
\end{itemize}

\section{PhaseGAN}

\subsection{Overview}
\begin{figure*}[t]
\centering
\includegraphics[width=\textwidth]{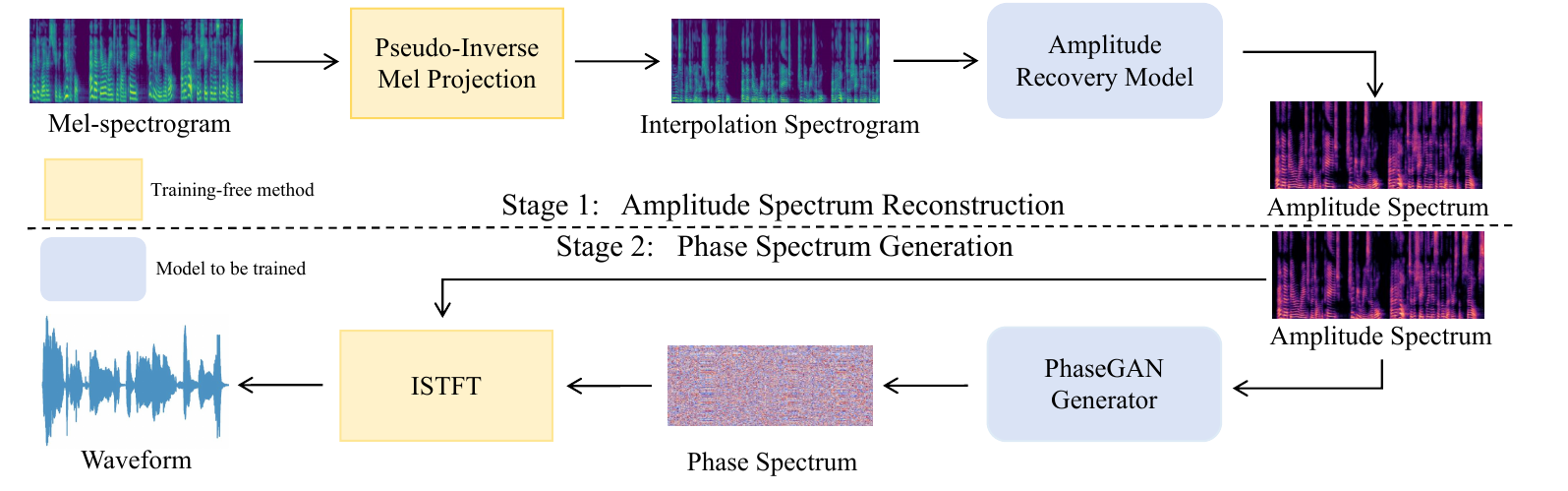}
\caption{PhaseGAN Inference Pipeline.}
\label{figure1}
\end{figure*}

As shown in Figure~\ref{figure1}, the vocoder's inference pipeline comprises four steps: First, the Mel-spectrogram is transformed to match the spectral dimension of the original amplitude spectrum via Mel-filter bank pseudo-inversion \cite{freeV3}. Second, an ICCRN reconstructs the amplitude spectrum. Third, the generator synthesizes the phase spectrum conditioned on this amplitude estimate. Finally, we use amplitude and phase spectra to synthesize the audio waveform by inverse Fourier transformation.

\subsection{ICCRN}

The detailed architecture of ICCRN can be found in our previous work \cite{iccrn}. As an in-place model, ICCRN avoids any frequency downsampling or upsampling operations. This mitigates information loss, prevents the introduction of extraneous information, and directly circumvents the artifact problem caused by upsampling operations. Consequently, the frequency dimension of the model remains constant (f=513), and the number of output channels is identical across all convolutional layers. To further analyze time-frequency (T-F) domain features, the model applies cepstral analysis. ICCRN also utilizes LSTM \cite{lstm} along frequency and time axis for dual-path modeling \cite{lstm2}. To ensure the causality of the model, LSTM along time axis is unidirectional, which means no future frame is used. This cepstral analysis capability enables ICCRN to examine the relationship between amplitude and phase information from a harmonic structure perspective; As the core neural network architecture, ICCRN is used in the reconstruction of both the amplitude spectrum and the phase spectrum. Given the distinct mathematical properties and value ranges of amplitude and phase, the corresponding recovery and reconstruction networks differ primarily in the design of their output layers. Further details will be provided in later sections.

\subsection{Amplitude Spectrum Reconstruction}
\subsubsection{Network}

An interpolation is performed on the 80-dimensional Mel-spectrogram using the pseudo-inverse operation of the Mel-scale filterbank. This process restores the features to their original frequency domain dimensionality (f = 513). The interpolated spectrum is subsequently processed by a amplitude recovery model, which is an ICCRN network. Key architectural modifications include: unifying the number of output channels for all convolutional layers to 10 (c = 10), and adopting the Softplus activation function for the final output layer. The model takes the interpolation spectrogram as input and generates the optimized amplitude spectrum as output.

\subsubsection{Training Loss}

The mean square error (MSE) loss is employed for amplitude spectrum estimation. Seen in Equation \eqref{eq1}:
\begin{equation}
{\mathcal{L}}_{\text{MSE}} = \frac{1}{T \cdot F} \sum_{t=1}^{T} \sum_{f=1}^{F} \left( |\hat{X}(t,f)| - |X(t,f)| \right)^{2} \tag{1}
\label{eq1}
\end{equation}
where \(|{X}(t,f)|\) and \(|\hat{X}(t,f)|\) denote the unit of target and estimate amplitude spectrum at  frame \textit{t} and frequency \textit{f}, respectively.

\subsection{Phase Spectrum Generation}
\subsubsection{Generator}

The generator in proposed PhaseGAN similarly employs the ICCRN architecture. The distinction lies in standardizing the number of output channels across all convolutional layers to 20 (c = 20). Regarding the output layer design, we draw upon the APNet \cite{apnet} architecture. The model employs a parallel estimation architecture whose design integrates two linear convolutional layers and the phase formula \(\Phi\). This approach is inspired by deriving phase spectra from the real and imaginary components of complex spectra. The pseudo-imaginary component \(\hat{I} \in \mathbb{R}^{T \times F}\) and the pseudo-real component \(\hat{R} \in \mathbb{R}^{T \times F}\) are the outputs of the two parallel layers. \(\Phi\) then calculates the phase spectrum \(\hat{\theta}\). For \(\forall R \in \mathbb{R}\) and \(I \in \mathbb{R}\), we define the \(\Phi\) by Equation  \eqref{eq2}:
\begin{equation}
\Phi \left( {R,I}\right)  = \arctan \left( \frac{I}{R}\right)  - \frac{\pi }{2} \cdot  {\operatorname{Sgn}}^{ * }\left( I\right)  \cdot  \left\lbrack  {{\operatorname{Sgn}}^{ * }\left( R\right)-1}\right\rbrack\tag{2}
\label{eq2}
\end{equation}
and \(\Phi(0, 0) = 0\). When \(x \geq 0, \operatorname{Sgn}^*(x)\) equals 1; otherwise, it is -1. Formula \(\Phi\) strictly restricts the predicted phase to the principal value interval (-$\pi$,$\pi$] for direct wrapped phase prediction.

\subsubsection{Discriminator}

The discriminator in PhaseGAN employs a multi-scale architecture (inspired by MelGAN), which takes time-domain speech waveforms as input. It consists of multiple parallel discriminator modules with identical structures, each comprising a three-layer 1D convolutional stack. These modules perform downsampling on input speech features via multi-layer convolutional operations, and their outputs directly drives the adversarial training process for the generator.

\subsubsection{Training Loss}

To train the model of phase generation, we use the R3GAN \cite{r3gan}, which significantly ehances training stability by integrating the relativistic paried GAN loss with zero centered gradient penalty mechanism \cite{zgp}. Experimental results of R3GAN demonstrate that even without the introduction of any specific training techniques, this generative method still exhibits favorable convergence performance and training stability. Due to the high sensitivity of phase information, conventional generative methods often face challenges in achieving stable convergence; therefore, the proposed approach provides a valuable reference for generative phase reconstruction.

The generator’s loss function is defined by \eqref{eq3} :
\begin{equation}
\mathcal{L}_{G} = \alpha \cdot \mathcal{L}_{\text{mrstft}}(\bm{\hat{y}},\bm{y}) + \beta \cdot \mathcal{L}_{\text{Adv}}(D(\bm{\hat{y}}),D(\bm{y}))
\tag{3}
\label{eq3}
\end{equation}
where \(G\) denotes the generator, \(D\) represents the discriminator, \(y\) signifies the real audio sample, and \(\hat{y}\) corresponds to the synthesized speech sample. $\alpha$ and $\beta$ denote the weighting coefficients.

The generator loss consists of adversarial losses and is defined as Equation \eqref{eq4}:
\begin{equation}
\mathcal{L}_{Adv}( \bm{y},\bm{\hat{y}}) =\mathbb{E}_{(\bm{y},\bm{\hat{y}})}\biggl[\operatorname{softplus}\left(-(\bm{y} -\bm{\hat{y}})\right)\biggr]\tag{4}
\label{eq4}
\end{equation}
where Softplus refers to the softplus activation function, which restricts values to be greater than zero.

To enhance the consistency between phase reconstruction and the recovered amplitude spectrum, a multi-resolution STFT loss is incorporated into the generator's objective function. This loss term indirectly constrains phase gradients, guiding the model to progressively reconstruct phase information that exhibits high correspondence with the amplitude spectrum during training. The mathematical formulation of $\mathcal{L}_{mrstft}$ is defined by \eqref{eq5}:
\begin{equation}
{\mathcal{L}}_{mrstft}\left( {\hat{y},y}\right)  = \frac{1}{K}\mathop{\sum }\limits_{{k = 1}}^{K}\left\lbrack  {{\mathcal{L}}_{\text{mag }}^{\left( k\right) }\left( {\hat{y},y}\right)  + {\mathcal{L}}_{sc}^{\left( k\right) }\left( {\hat{y},y}\right) }\right\rbrack\tag{5}
\label{eq5}
\end{equation}
where $K=3$ indicates the number of distinct resolution configurations. ${\mathcal{L}}_{\text{mag }}^{\left( k\right) }$ is the amplitude spectral loss and ${\mathcal{L}}_{sc}^{\left( k\right) }$ is the spectral convergence loss which defined by \eqref{eq6}--\eqref{eq7}:
\begin{equation}
{\mathcal{L}}_{\text{mag }}^{\left( k\right) }\left( {\hat{y},y}\right)  = {\begin{Vmatrix}\left| STF{T}^{\left( k\right) }\left( \hat{y}\right) \right|  - \left| STF{T}^{\left( k\right) }\left( y\right) \right| \end{Vmatrix}}_{1}  \tag{6}
\label{eq6}
\end{equation}
\begin{equation}
{\mathcal{L}}_{sc}^{\left( k\right) }\left( {\hat{y},y}\right)  = \frac{{\begin{Vmatrix}\left| STF{T}^{\left( k\right) }\left( \hat{y}\right) \right|  - \left| STF{T}^{\left( k\right) }\left( y\right) \right| \end{Vmatrix}}_{F}}{{\begin{Vmatrix}\left| STF{T}^{\left( k\right) }\left( y\right) \right| \end{Vmatrix}}_{F} + \epsilon }\tag{7}
\label{eq7}
\end{equation} 

where \(\Vert \cdot \Vert_1\) denotes the L1-norm and \(\Vert \cdot \Vert_{\text{F}}\) denotes the Frobenius norm, which is a matrix norm that quantifies the overall scale of a matrix, with \(\epsilon\) representing an infinitesimal positive constant to ensure numerical stability.

The discriminator loss is composed of adversarial losses and a zero-centered gradient penalty term. The discriminator's loss function is defined as follows \eqref{eq8}:
\begin{equation}
\begin{aligned}
\mathcal{L}_{D} 
&= \mathbb{E}_{(\bm{y},\bm{\hat{y}})}
\biggl[ \operatorname{softplus}\Bigl( -\bigl( D(\bm{y}) - D(\bm{\hat{y}}) \bigr) \Bigr) \biggr] \\
&\quad + \frac{\gamma}{2} \biggl( 
    \underbrace{\mathbb{E}_{\bm{y}} \Bigl[ \lVert \nabla_{\bm{y}} D(\bm{y}) \rVert^{2} \Bigr]}_{R_{1}}
    + \underbrace{\mathbb{E}_{\bm{\hat{y}}} \Bigl[ \lVert \nabla_{\bm{\hat{y}}} D(\bm{\hat{y}}) \rVert^{2} \Bigr]}_{R_{2}}
\biggr)
\end{aligned}
\label{eq8}
\tag{8}
\end{equation}
where $\gamma$ denotes the dynamic decay coefficient (characterized by a cosine scheduler) \cite{SGDR}, $\nabla_{\!{y}} D({y})$ represents the discriminator's gradient with respect to the real sample input, $\nabla_{\!{\hat{y}}} D({\hat{y}})$ corresponds to the discriminator's gradient with respect to the generated sample input, and $\{R_1, R_2\}$ indicate zero-centered gradient penalties \cite{zgp2}.

\section{Experimental Setup}
\subsection{Datasets}

To train and evaluate the performance of the proposed PhaseGAN, we employ three datasets.
\subsubsection{Single speaker speech synthesis}
The LJSpeech dataset \cite{lj_speech} was used for a single speaker experiment\footnote{https://keithito.com/LJ-Speech-Dataset/}. This publicly available benchmark dataset comprises 13,100 audio clips recorded by a single female English speaker, about 24 hours. The audio clips are sampled at 22.05 kHz with 16 bit. We partitioned the dataset into training, validation, and test sets according to Guideline from the open-source VITS
repository\footnote{{https://github.com/jaywalnut310/vits/tree/
main/filelists}}. This dataset is commonly employed for performance comparison among different neural vocoders.
\subsubsection{Unseen speaker speech synthesis}
VCTK \cite{vctk}, a multi-speaker dataset\footnote{{https://datashare.ed.ac.uk/handle/10283/3443}}, was used to evaluate the generalization ability for unseen speakers outside the training data. It comprises approximately 44,200 short audio utterances from 109 native English speakers with diverse accents, totaling 44 hours. The original audio recordings were sampled at 44 kHz and downsampled to 22.05 kHz in our experiment. To evaluate PhaseGAN's generalization for Mel-spectrogram inversion on unseen speakers, we randomly selected two speakers ( \(\approx\) 500 utterances) for testing using models trained exclusively on the single-speaker LJSpeech dataset (without fine-tuning). PhaseGAN's inversion results were compared against those from HiFi-GAN and FreeV, which were also trained solely on LJSpeech.

\subsubsection{Singing Voice speech synthesis}

To further validate generalization capability, we also evaluate our method on singing voice data that was not included in the training set. We utilized the Opencpop dataset \cite{opencpop}–a Mandarin singing corpus containing 100 pop songs performed by a trained female vocalist\footnote{{https://wenet.org.cn/opencpop/}}. Original audio files were recorded at 44.1 kHz sampling rate. For experimental consistency, we downsampled all audio to 22.05 kHz, removed silent segments using Voice Activity Detection (VAD) \cite{VAD}, and segmented utterances into 10-second clips for model evaluation. Additionally, for fair comparison, we conducted a comparative analysis between the synthesized outputs of our proposed model and those of HiFi-GAN and FreeV. All models were evaluated without any fine-tuning.

\subsection{Baseline}
To compare vocoder performance, this study chose several representative models for benchmark, including HiFi-GAN \cite{hifigan}, iSTFTNet \cite{istftnet}, APNet \cite{apnet}, BigVGAN \cite{BIGVGAN}, APNet2 \cite{apnet2}, Vocos \cite{vocos} and FreeV \cite{freeV3}. All these models provide officially released pretrained models based on the LJSpeech dataset. However, inconsistencies exist in their official data splits (specifically, the definitions of the training, validation, and test sets). To ensure a fair comparison, all models were uniformly retrained using the open-source data splits provided by the VITS codebase. All baseline models strictly aligned their parameters with the officially released optimal versions.

\subsection{Evaluation Metrics}
\subsubsection{Subjective evaluation}
To evaluate the audio quality, we crowd-sourced 5-scale MOS tests via Amazon Mechanical Turk. Participants rated 
speech samples on a scale from 1 (’poor-completely unnatural speech’) to 5 (’excellent - completely natural speech’).
The MOS scores were recorded with 95\% confidence intervals (CI). Raters listened to the test samples randomly, where they were allowed to evaluate each audio sample once. All audio clips were
normalized to prevent the influence of audio volume differences on the raters. In the MOS test, 20 listeners participated (for the English test, it was native English speakers; for the singing test, it was 5 native Mandarin speakers), which ensured the reliability.  All quality assessments in Section 4 were conducted in this manner, and were not sourced from other papers.

\subsubsection{objective evaluation}
This study employs more objective metrics for thorough evaluation, including the F1 score for voiced/unvoiced classification (V/UV F1, optimal: 1.0), Periodicity error (optimal: 0.0), Pitch RMSE (optimal: 0.0), F0 RMSE (optimal: 0.0), UTMOS (optimal: 5.0) \cite{utmos}, wide-band perceptual evaluation speech quality (WB-PESQ, optimal: 4.5) \cite{pesq}, short time objective intelligence  (STOI, optimal: 1.0) \cite{stoi}, mel-cepstrum distortion (MCD, optimal: 0.0). Specifically, Pitch RMSE and F0 RMSE quantify pitch accuracy, while UTMOS and WB-PESQ target perceptual quality. For spectral fidelity, Periodicity error and MCD evaluate the synthesized speech, whereas STOI and V/UV F1 gauge speech intelligibility.

\subsection{Training Setup}
We trained all models on the LJSpeech dataset. For our proposed two-stage framework, the Amplitude recovery model (Stage 1) was trained for 1 million steps with 16 minibatch size, while the phase generation model (Stage 2) underwent 1.5 million steps with 16 minibatch size. The baseline model's training steps and minibatch size match those of our proposed method. All models shared the same experimental setup. For feature extraction, a 1024-point Fast Fourier Transform (FFT) was performed, utilizing a Hanning window of length 1024 and a hop size of 256 samples. We used 80 mel-frequency bands with a frequency cutoff of 16 kHz. The sampling rate was 22,050 Hz. Optimization employed AdamW \cite{adamw} with an initial learning rate of 0.0002, optimizer parameters ($\beta_{1}$, $\beta_{2}$) set to (0.8, 0.99), and exponential learning rate decay applied with a decay factor of 0.999. Each audio sample was processed via random window cropping, resulting in segments of 16,384 samples (approximately 0.74 seconds). For the multi-resolution STFT loss (${\mathcal{L}}_{mrstft}$), this study adopts three distinct time-frequency resolution configurations (window-hop-FFT size): (400, 80, 512), (800, 200, 1024) and (1600, 400, 2048). The weighting parameters $\alpha$ and $\beta$ in the generator's loss function are empirically configured at 0.8 and 0.2, respectively.

This study trained two variants of PhaseGAN: standard and lightweight version, named \textbf{PhaseGAN} and \textbf{PhaseGAN-s}. In the PhaseGAN (standard version), the Amplitude Recovery Module employs a 10-channel ICCRN model, while the Generator employs a 20-channel ICCRN model. In the PhaseGAN-s (lightweight version), the Amplitude Recovery Module and Generator employ 3-channel and 8-channel ICCRN models, respectively. The number of Cepstral Frequency Blocks (CFB) layers were uniformly set to 5 across all models.

\section{Experimental Results}

\subsection{Single-Speaker Expressiveness}

Table~\ref{tab1} presents a comparison between the proposed model and baseline methods. Based on the comprehensive comparison in Table 1 of vocoder performance metrics, with only 1.6M parameters, PhaseGAN achieves SOTA in all perceptual quality metrics, like highest UTMOS 4.238, the best subjective human-rated quality, lowest MCD 2.108 indicating superior mel-cepstral fidelity and naturalness. Lowest periodicity 0.088 for PhaseGAN and 0.112 for PhaseGAN-s shows that superior phase estimation reduces "metallic" artifacts common in GAN vocoders. It can also be observed that PhaseGAN obtain lowest Pitch-RMSE 17.926 and F0-RMSE 35.216, which are crucial for natural prosody. PhaseGAN-s offers near-SOTA quality reducing the number of parameters by over 90\%, which is suitable for edge devices. At the same time, our subjective MOS (Media Opinion Score) index is also the best.
It is noteworthy that the UTMOS score of authentic speech reaches 4.37, demonstrating the high fidelity achieved by our method.  We refer to \cite{nvase} for the reproduction of the baseline model.

\begin{table*}[t]
\centering
\footnotesize
\caption{Objective evaluation results on the LJSpeech single-speaker speech synthesis test set. The last Five rows present results from ablation studies on PhaseGAN, with the best performance highlighted in bold.}
\label{tab:model_comparison}
\begin{threeparttable}
\hspace*{-0.3cm}
\begin{adjustbox}
{width=1.00\textwidth}
\begin{tabular}{L *{12}{c}}
\toprule
\bfseries Model & \bfseries Domain & \bfseries \# Mac & \bfseries \# Params & \bfseries WB-PESQ$\uparrow$ & \bfseries STOI$\uparrow$ & \bfseries MCD$\downarrow$ & \bfseries V/UV F1$\uparrow$ & \bfseries Periodicity$\downarrow$ & \bfseries Pitch-RMSE$\downarrow$ & \bfseries F0-RMSE$\downarrow$ & \bfseries UTMOS$\uparrow$ & \bfseries MOS$\uparrow$ \\
\midrule
HiFiGAN & T & 29.44G & 13.94M & 3.574 & 0.931 & 3.641 & 0.953 & 0.125 & 32.279 & 36.232 & 4.219 & 4.233 (±0.03) \\
\addlinespace[0.04cm] 
iSTFTNet & T & 20.75G & 13.26M & 3.535 & 0.932 & 3.625 & 0.951 & 0.124 & 35.096 & 37.436 & 4.236 & 4.143 (±0.07) \\
\addlinespace[0.04cm]
BigVGAN & T & 27.83G & 14.01M & 3.492(3.519) & 0.965 & 3.168 & 0.956(0.945) & 0.127(0.128) & 30.842 & 37.142& 4.137 & 3.982 (±0.04) \\
\addlinespace[0.04cm]
APNet & T-F & 5.99G & 72.19M & 3.391 & 0.968 & 3.285 & 0.949 & 0.142 & 21.252 & 39.739/(19.83) & 3.177 & 3.783 (±0.04) \\
\addlinespace[0.04cm]
Vocos & T-F & 1.11G & 13.35M & 3.522/(3.70) & 0.973 & 2.672 & 0.957/(0.958) & 0.115/(0.101) & 28.185 & 36.561 & 3.973/(3.734) & 4.112 (±0.04)\\
\addlinespace[0.04cm]
APNet2 & T-F & 2.61G & 31.43M & 3.492 & 0.971 & 2.829/(2.078) & 0.962 & 0.108 & 26.034 & 40.046/(44.33) & 3.938 & 3.874 (±0.05) \\
\addlinespace[0.04cm]
FreeV & T-F & 1.51G & 18.22M & 3.593/(3.431) & 0.975/(0.967) & 2.752/(3.112) & 0.962/(0.956) & 0.106/(0.118) & 24.418 & 39.087/(26.40) & 4.015 & 4.017 (±0.06)\\
\addlinespace[0.04cm]
PhaseGAN-s  & T-F & 1.05G & 0.54M & 3.672 & 0.981 & 2.483 & 0.952 & 0.112 & 20.522 & 36.623 & 4.019 & 3.986 (±0.05) \\
\addlinespace[0.04cm]
PhaseGAN & T-F & 6.42G & 1.63M & \textbf{3.926} & \textbf{0.988} & \textbf{2.108} & \textbf{0.969} & \textbf{0.088} & \textbf{17.926} & \textbf{35.216} & \textbf{4.238} & \textbf{4.256 (±0.05)} \\
\cmidrule{1-13} 
\addlinespace[0.04cm] 
\multicolumn{1}{l}{-w/o mrstft.} & T-F & 6.42G & 1.63M & 3.482 & 0.974 & 2.682 & 0.956 & 0.112 & 24.572 & 41.072 & 3.875 & 3.586 (±0.07) \\
\addlinespace[0.04cm]
\multicolumn{1}{l}{-w/o zgp.} & T-F & 6.42G & 1.63M & 3.091 & 0.966 & 3.062 & 0.952 & 0.126 & 34.221 & 49.523 & 3.012 & 3.014 (±0.04) \\
\addlinespace[0.04cm]
\multicolumn{1}{l}{-w/o adv.} & T-F & 6.42G & 1.63M & 3.622 & 0.956 & 2.345 & 0.943 & 0.122 & 23.426 & 38.782 & 3.483 & 3.743 (±0.05) \\
\addlinespace[0.04cm]
\multicolumn{1}{l}{-w/o freq.} & T-F & 6.32G & 1.60M & 3.582 & 0.981 & 2.463 & 0.956 & 0.098 & 21.516 & 37.385 & 3.958 & 3.686 (±0.04) \\
\addlinespace[0.04cm]
\multicolumn{1}{l}{-w/o ceps.} & T-F & 5.78G & 1.60M & 3.695 & 0.983 & 2.382 & 0.962 & 0.102 & 21.453 & 39.721 & 3.534 & 3.667 (±0.04) \\
\bottomrule
\end{tabular}
\end{adjustbox}
\begin{tablenotes}
    \item \footnotesize{\textit{Note:} Values in parenthese are copied from the original papers.}
\end{tablenotes}
\end{threeparttable} 
\label{tab1}
\end{table*}

\subsection{Ablation Study}

To validate the effectiveness of key components, we conducted ablation studies on the core loss functions of the training strategy and the essential structure of the ICCRN model. The results are presented in Table~\ref{tab1}. Removal of Multi-resolution STFT Loss (w/o mrstft): Eliminating this term from the generator's loss function resulted in significant degradation across all performance metrics. This decline is attributed to reduced fitting accuracy of phase and amplitude information, consequently impairing synthesized speech quality (audible artifacts can be observed in the provided audio samples). Removal of Zero-Centered Gradient Penalty (w/o zgp): Excluding this penalty term from the discriminator's loss function diminished the model’s convergence capability, leading to substantial deterioration in speech synthesis performance metrics. Removal of Adversarial Loss (w/o adv): the accompanying audio samples (refer to provided examples) revealed pronounced mechanical artifacts in the synthesized speech, significantly degrading perceptual quality. This perceptual decline is further corroborated by lower UTMOS scores. Removal of ICCRN Core Modules: The ablation of the freq module (denoting the LN → Conv3x1 module within the CFB) and the ceps module (denoting the Ceps Unit) also resulted in performance degradation. This outcome confirms the critical role both modules play in the task performance of the ICCRN architecture.

\subsection{Generalization to Unseen Speakers}

To evaluate the generalization capability of the model, this study conducted tests on the VCTK dataset using the model without fine-tuning. As shown in Table~\ref{tab2}, the evaluation results demonstrate that the proposed method exhibits robust generalization ability to unseen speakers.The synthesized speech generated by our method demonstrates significantly superior performance across multiple metrics compared to HiFi-GAN and FreeV. This indicates that the phase generation process underlying the method primarily depends on amplitude spectrum information, rather than speaker-specific characteristics. Experimental results indicate that in subjective listening tests, the speech synthesized by our method for unseen speakers outperforms the baseline models in both clarity and naturalness. 

\begin{table*}[t] 
\centering
\small 
\caption{Objective evaluation results on the VCTK unseen-speaker speech synthesis test set (without fine-tuning). The best-performing results are highlighted in bold.}
\label{tab:model_comparison2}
\begin{adjustbox}{width=0.95\textwidth, center}
\begin{tabular}{L *{10}{c}} 
\toprule
\bfseries Model & \bfseries Domain & \bfseries WB-PESQ$\uparrow$ & \bfseries STOI$\uparrow$ & \bfseries MCD$\downarrow$ & \bfseries V/UV F1$\uparrow$ & \bfseries Periodicity$\downarrow$ & \bfseries Pitch-RMSE$\downarrow$ & \bfseries F0-RMSE$\downarrow$ & \bfseries UTMOS$\uparrow$ & \bfseries MOS$\uparrow$ \\
\midrule
HiFiGAN & T & 2.578 & 0.929 & 3.212 & 0.887 & 0.208 & 37.109 & 39.542 & 3.232 & 3.886 (±0.06) \\
\addlinespace[0.04cm] 
FreeV & T-F & 2.414 & 0.946 & 3.802 & 0.897 & 0.199 & 45.382 & 48.172 & 3.131 & 3.764 (±0.05) \\
\addlinespace[0.04cm] 
PhaseGAN-s & T-F & 3.041 & 0.949 & 2.883 & 0.936 & 0.134 & 18.645 & 37.452 & 3.275 & 3.904 (±0.07) \\
\addlinespace[0.04cm] 
PhaseGAN & T-F & \textbf{3.199} & \textbf{0.963} & \textbf{2.595} & \textbf{0.941} & \textbf{0.127} & \textbf{16.026} & \textbf{36.772} & \textbf{3.357} & \textbf{4.142 (±0.07)} \\
\bottomrule
\end{tabular}
\label{tab2}
\end{adjustbox}
\end{table*}

\subsection{Generalization to Singing Voice}

To evaluate the model's cross-lingual singing voice generalization capability, we conduct zero-shot testing by directly transferring the model trained on the English speech dataset (LJSpeech) to the Mandarin singing voice dataset (Opencpop). As indicated in Table~\ref{tab3}, despite the dual disparities in linguistic properties (English vs. Mandarin), articulation modes (speech vs. singing), and acoustic distributions, the proposed method demonstrates exceptional cross-domain robustness. According to the provided audio sample, it can be known that, baseline models exhibit noticeable underperformance on the cross-domain singing dataset, producing rather blurred harmonic structures. In contrast, the proposed method reconstructs results that are comparable in quality to the reference audio, with clear and complete harmonics. For specific auditory examples, please refer to the provided audio samples.

\begin{table*}[t] 
\centering
\small 
\caption{Objective evaluation results on the Opencpop singing voice dataset (without fine-tuning). The best-performing results are highlighted in bold.}
\label{tab:model_comparison3}
\begin{adjustbox}{width=0.95\textwidth, center}
\begin{tabular}{l c c c c c c c c c}
\toprule
\bfseries Model & \bfseries Domain & \bfseries WB-PESQ$\uparrow$ & \bfseries STOI$\uparrow$ & \bfseries MCD$\downarrow$ & \bfseries V/UV F1$\uparrow$ & \bfseries Periodicity$\downarrow$ & \bfseries Pitch-RMSE$\downarrow$ & \bfseries F0-RMSE$\downarrow$ & \bfseries MOS$\uparrow$ \\
\midrule
HiFiGAN & T & 2.301 & 0.821 & 3.532 & 0.942 & 0.268 & 25.193 & 39.251 & 4.012 (±0.04) \\
\addlinespace[0.04cm] 
FreeV & T-F & 2.952 & 0.842 & 3.011 & 0.944 & 0.357 & 18.312 & 43.412 & 3.956 (±0.07) \\
\addlinespace[0.04cm] 
PhaseGAN-s & T-F & 3.613 & 0.892 & 2.162 & 0.931 & 0.257 & 13.451 & 41.399 & 4.132 (±0.05) \\
\addlinespace[0.04cm] 
PhaseGAN & T-F & \textbf{3.865} & \textbf{0.922} & \textbf{1.512} & \textbf{0.952} & \textbf{0.211} & \textbf{11.961} & \textbf{39.112} & \textbf{4.356 (±0.07)} \\
\bottomrule
\end{tabular}
\label{tab3}
\end{adjustbox}
\end{table*}

\subsection{End-to-End Speech Synthesis}

To validate the effectiveness of the proposed model in an end-to-end speech synthesis pipeline, we designed an experimental study. The pipeline consists of two stages: text-to-mel-spectrogram conversion and mel-spectrogram-to-waveform synthesis. First, mel-spectrograms were generated using FastSpeech2, with no structural modifications applied; instead, its widely-used public implementation and pre-trained weights were directly adopted. Subsequently, the generated mel-spectrograms were used as conditional input to the second-stage vocoders, which included the standard and lightweight versions of our proposed model, as well as the baseline systems HiFiGAN and FreeV, for waveform synthesis. The validation data consisted of the test set corresponding to LJSpeech, and all vocoders were used in their pre-trained versions without any fine-tuning, directly synthesizing speech from the mel-spectrograms.

Table~\ref{tab4} presents the results of each model in terms of the UTMOS objective score and subjective mean opinion score (MOS). It can be observed that the proposed method not only demonstrates good generalization ability on unseen speaker and singing datasets, but also shows high compatibility with acoustic model-driven speech synthesis application scenarios. According to the provided audio sample, it can be known that, the proposed method produces virtually no artifacts in the high-frequency range, while reconstructing clearer and more complete harmonics in the low-frequency region compared to the baseline model.

\begin{table}[t]
\centering
\small
\caption{The end-to-end speech synthesis quality comparison of\\
different vocoders without any fine-tuning on the LJSpeech
test set. The best-performing results are highlighted in bold.}
\label{tab:model_comparison3}

\begin{tabular}{l c c c}
\toprule
\bfseries Model & \bfseries Domain & \bfseries UTMOS$\uparrow$ & \bfseries MOS$\uparrow$\\
\midrule
HiFiGAN       & T     & 4.132 & 4.145 (±0.02) \\
FreeV         & T--F  & 3.952 & 4.052 (±0.04) \\
PhaseGAN-s    & T--F  & 3.978 & 3.985 (±0.07) \\
PhaseGAN      & T--F  & \textbf{4.245} & \textbf{4.189 (±0.06)} \\
\bottomrule
\end{tabular}
\label{tab4}
\vspace{1.5em}
\end{table}

\section{Conclusion}

This study proposes PhaseGAN, a model capable of synthesizing high-quality, artifact-free speech with ultra-low parameter counts. Experimental results demonstrate its superior synthesis quality over state-of-the-art publicly available models. Inspired by the R3GAN framework from computer vision, we innovatively introduce a phase generation mechanism that effectively addresses the longstanding phase reconstruction challenge in vocoders. To the best of our knowledge, this represents the first successful demonstration of phase-related problem solving via generative adversarial networks. Additionally, PhaseGAN exhibits exceptional generalization capabilities. While maintaining computational complexity comparable to the current optimal public model (FreeV), our lightweight implementation achieves a substantial reduction in parameters while delivering superior audio fidelity. This breakthrough provides new pathways for edge-device natural speech synthesis requiring low latency and minimal memory footprint. And, this work demonstrates a strong potential for industrial scene deployment.

\section{Future Work}

We will further validate the effectiveness of the proposed method on more diverse datasets, encompassing a wider variety of speakers and languages. In our future work, we will validate our phase reconstruction method on various models to further demonstrate its universality. Additionally, the method of phase reconstruction will be extended to a broader range of application domains, such as codec, speech enhancement, and bandwidth extension.

%
%
%
%

\end{document}